\documentclass[twocolumn]{aastex631}

\usepackage[normalem]{ulem}
\usepackage{color} 

\def\hal{{\rm H}\alpha}
\def\hbe{{\rm H}\beta}

\received{\today}

\begin{document}

\title{RNAAS \\
  JWST spectra are consistent with the edge-on star-forming galaxy scenario \\
  for the ''runaway supermassive black hole''
} 

\correspondingauthor{J. S\'anchez Almeida} 
\email{jos@iac.es}

\author[0000-0003-1123-6003]{Jorge S\'anchez Almeida}  \affil{All authors contributed equally} \affil{Instituto de Astrof\'\i sica de Canarias, La Laguna, Tenerife, E-38200, Spain} \affil{Departamento de Astrof\'\i sica, Universidad de La Laguna}

\author[0000-0001-8647-2874]{Ignacio Trujillo} \affil{All authors contributed equally}
 \affil{Instituto de Astrof\'\i sica de Canarias, La Laguna, Tenerife, E-38200, Spain} \affil{Departamento de Astrof\'\i sica, Universidad de La Laguna} 
\author[0000-0001-6444-9307]{Sebasti\'an F. S\'anchez}\affil{All authors contributed equally}
 \affil{Instituto de Astronom\'\i a, Universidad Nacional Aut\'onoma de M\'exico, A.P. 106, Ensenada 22800, BC, Mexico} \affil{Instituto de Astrof\'\i sica de Canarias, La Laguna, Tenerife, E-38200, Spain} \affil{Departamento de Astrof\'\i sica, Universidad de La Laguna}

\author[0000-0001-7847-0393]{Mireia Montes}\affil{All authors contributed equally}
 \affil{Institute of Space Sciences (ICE, CSIC), Campus UAB, Carrer de Can Magrans, s/n, 08193 Barcelona, Spain}


\begin{abstract}
The linear structure reported by van Dokkum et al. (2023) has been proposed as either a massive stellar wake produced by a runaway supermassive black hole (SMBH) or a bulgeless edge-on galaxy. New JWST/NIRSpec IFU observations target the tip of the structure, where a SMBH would produce a bow shock, whereas a normal galaxy would host an HII region. Using standard BPT diagrams ([OIII]5007/$\hbe$ vs [NII]6583/$\hal$ and [OIII]5007/$\hbe$ vs [OII]6716,6731/$\hal$), we find that the line ratios at the tip fall on the locus of low-metallicity low-extinction HII regions. This region does not overlap with loci typical of shocks in merging galaxies. Thus, these results are consistent with the interpretation that the linear structure is a star-forming galaxy, with the bright knot representing one of its HII regions.
\end{abstract}




\section{Rationale}

The linear structure ($\sim$45 kpc long, $\sim$1 kpc wide, at redshift $\sim$1) reported by \citet{2023ApJ...946L..50V} has attracted considerable attention because it was proposed to represent a massive stellar wake ($\sim 3\times 10^9 M_\odot$) produced as a supermassive black hole (SMBH), ejected from a nearby galaxy, plowed through a giant gas cloud. The ejection of SMBHs from galaxies has been theorized \citep[e.g.,][]{1974ApJ...190..253S}, but never observed on such scales. Unfortunately, this engaging explanation is far from clear-cut, as the mass, rotation, and morphology of the structure closely resemble those of a bulgeless edge-on galaxy in the rest-frame UV \citep{2023A&A...673L...9S}. The evidence for either scenario is discussed in a number of papers
\citep{
  2023A&A...673L...9S,
  2023A&A...678A.118S,
  2023ApJ...946L..50V,
  2023RNAAS...7...83V,
  2024RNAAS...8..150M}.
This Research Note is triggered by new JWST/NIRSpec IFU observations of the tip of the structure,  where the SMBH is expected to lie \citep{2025arXiv251204166V,https://doi.org/10.17909/8d8q-r813}. These new observations provide part of the optical rest-frame emission-line spectrum, for which many diagnostic diagrams have been developed over the years to distinguish emission produced by shocks from that produced by star formation. This diagnostic is relevant because the physical mechanism producing the emission differs between the two scenarios: a runaway SMBH would generate a bow shock, whereas a normal galaxy would have an HII region powered by star formation. Here, we analyze the emission-line ratios measured in the bright knot at the tip of the structure using the standard diagnostic tools.

\section{BPT diagrams}

\begin{figure*}[ht!] 
\centering
\includegraphics[width=0.9\linewidth]{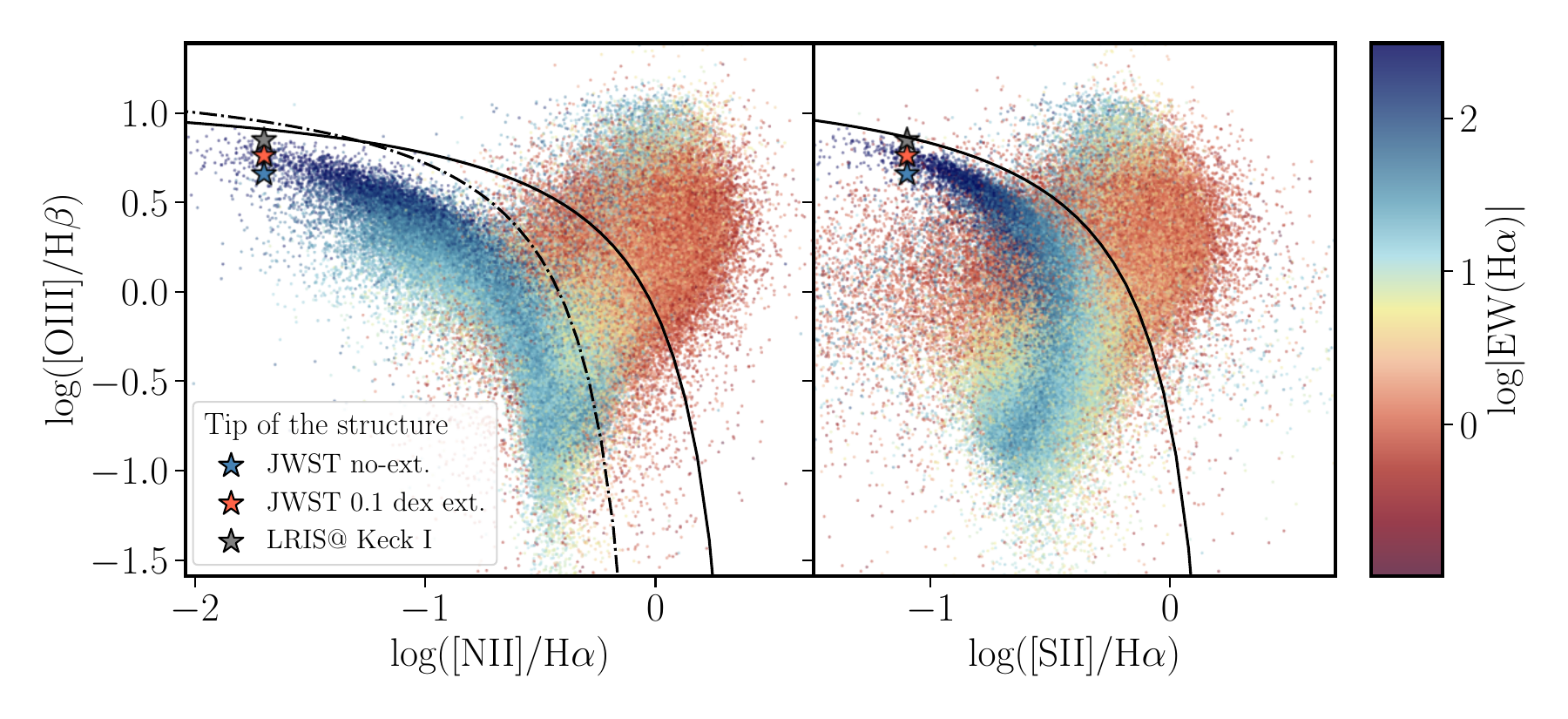}
\caption{BPT diagrams showing the location of the bright knot at the tip of the linear structure, proposed to be either a bulgeless edge-on galaxy \citep{2023A&A...673L...9S} or a runaway SMBH \citep{2023ApJ...946L..50V}. Left panel: [OIII]5007/$\hbe$ versus [NII]6583/$\hal$. Right panel: [OIII]5007/$\hbe$ versus [SII]6716,6731/$\hal$.  The colored dots are included to illustrate the line ratios observed in real galaxies. Each point corresponds to the central 3\arcsec\ region sampled by the SDSS fiber \citep[Sloan Digital Sky Survey;][]{2000AJ....120.1579Y} for approximately half a million local galaxies of all types, as selected by \citet{2025A&A...704A.145S}. The lines indicate the classical dividing curves defined by \citet[][solid lines]{2001ApJ...556..121K} and \citet[][dot–dashed line]{2003MNRAS.346.1055K}, below which star-forming regions are found. The dots are color-coded by the equivalent width of $\hal$ (shown in the side vertical bar), which, in the case of star-forming regions, serves as a proxy for the relative importance of the current starburst compared to past episodes. The symbols mark the tip of the structure for three different estimates of $\hbe$ (approach 1 in the text is labeled as LRIS@Keck I, whereas approach 2 corresponds to the orange and blue stars, with the third estimate overlapping the LRIS@Keck I value). In both BPT diagrams, the tip of the linear structure lies directly on the ridge of low-metallicity HII regions, actively forming stars at present \citep[e.g.,][]{2016ApJ...819..110S}. This part of the BPT diagrams does not overlap with where shocks in merging galaxies are typically found \citep[e.g.,][]{2011ApJ...734...87R,2014MNRAS.444.3894H}. Thus, the observed line ratios are fully consistent with the edge-on galaxy scenario.
}
\label{fig:fig2}
\end{figure*}
The standard way to distinguish emission-line spectra produced by HII regions, shocks, and AGNs (active galactic nuclei) is through the so-called BPT diagrams \citep[Baldwin-Phillips-Terlevich,][]{1981PASP...93....5B,2020ARA&A..58...99S}. Figure~\ref{fig:fig2} shows the two BPT diagrams that can be constructed using the emission lines from the new JWST spectra: [OIII]5007/$\hbe$ versus [NII]6583/$\hal$ (left panel) and [OIII]5007/$\hbe$ versus [SII]6716,6731/$\hal$ (right panel). The lines [OIII]5007, $\hal$ , [NII]6583, and [SII]6716,6731 are  within the  JWST band pass \citep[9700\,\AA\ -- 18900\,\AA ;][]{2025arXiv251204166V} but $H\beta$ (observed at 9547\,\AA ) lies outside.  Lacking $\hbe$ is not an obstacle, since (1) the ratio [OIII]5007/$\hbe$ was measured by \citet{2023ApJ...946L..50V} using the Low-Resolution Imaging Spectrometer (LRIS) on the Keck I telescope, and  (2) $\hbe$  can be inferred from the observed H$\alpha$ if they are produced by recombination. As we will show, the consistency of the two methods lends confidence to the result.
Concerning method (1), \citet[][Fig.~4]{2023ApJ...946L..50V} gives $\log({\rm [OIII]5007}/{\rm H}\beta)\simeq 0.85$ at the tip. As for approach (2), \citet[][Fig.~8]{2025arXiv251204166V} provide  ${\rm [OIII]5007}/\hal\simeq 1.6$, so that  $\log({\rm [OIII]5007}/{\rm H}\beta)=\log({\rm [OIII]5007}/{\rm H}\alpha)+\log({\rm H\alpha/H\beta})\simeq 0.66$ without extinction -- in this case $\hal/\hbe\simeq 2.86$ \citep[case~B recombination;][]{1974agn..book.....O}. The assumption of no extinction is consistent with the observed value for N2\,$\equiv$\,$\log({\rm [NII]}6583/\hal)\simeq -1.7$  (Fig.~8, at 62~kpc, on the bright knot), which pictures an HII region of low metallicity \citep[$\sim$20\,\% solar;][]{2004MNRAS.348L..59P}, characterized by low extinction \citep[e.g.,][]{2016ApJ...819..110S}. If a typical extinction coefficient in $\hbe$ between 0.1 and 0.2~dex is included in the estimate, it increases [OIII]5007 relative to $\hbe$, leaving  $\log({\rm [OIII]5007}/{\rm H}\beta)\simeq 0.76$\,--\,$0.86$.
These various estimates are shown in Fig.~\ref{fig:fig2} (star symbols), with  $\log({\rm [NII]}6583/\hal)$ given above and $\log({\rm [SII]}6716,6731/\hal)\simeq -1.1$ obtained from  \citet[][Fig.~8]{2025arXiv251204166V}.

\section{Result}
Figure 1 shows that the tip of the linear structure lies on the BPT diagrams in the locus of low-metallicity HII regions. This area does not overlap with the loci typically occupied by shocks in merging galaxies or outflows \citep[e.g.,][]{2011ApJ...734...87R,2014MNRAS.444.3894H,2021RMxAA..57....3S}. Thus,  the observed line ratios are consistent with  the interpretation that the linear structure is a galaxy, with the bright knot corresponding to one of its HII regions.
\newpage

\begin{acknowledgments}
JSA acknowledges financial support from the Spanish Ministry of Science and Innovation, project PID2022-136598NB-C31 (ESTALLIDOS8) and JSA from the EU UNDARK project (project number 101159929). 
IT acknowledges support from the State Research Agency (AEI-MCINN) of the Spanish Ministry of Science and Innovation under the grant PID2022-140869NB-I00 and IAC project P/302302, financed by the Ministry of Science and Innovation, through the State Budget and by the Canary Islands Department of Economy, Knowledge, and Employment, through the Regional Budget of the Autonomous Community. We also acknowledge support from the European Union through the following grants: "UNDARK" and "Excellence in Galaxies - Twinning the IAC" of the EU Horizon Europe Widening Actions programs (project numbers 101159929 and 101158446). Funding for this work/research was provided by the European Union (MSCA EDUCADO, GA 101119830).
SFS  thanks the support by UNAM PASPA -- DGAPA and the SECIHTI CBF--2025--I--236 project.
MM acknowledges support from grant RYC2022-036949-I financed by the MICIU/AEI/10.13039/501100011033 and by ESF+, grant CNS2024-154592 financed by MICIU/AEI/10.13039/501100011033 and program Unidad de Excelencia Mar\'{i}a de Maeztu CEX2020-001058-M.
\end{acknowledgments}

%

\end{document}